%% file: FVP-QAOA.tex
\documentclass[reprint,aps,prx,superscriptaddress,twocolumn,longbibliography]{revtex4-2}

\usepackage[utf8]{inputenc}
\usepackage[T1]{fontenc}
\usepackage{amsmath,amsfonts,amssymb,amsthm,amstext,bm}
\usepackage{graphicx}
\usepackage{xcolor}
\usepackage{tikz}
\usepackage{dcolumn}
\usepackage{booktabs}
\usepackage{subcaption}
\usepackage[colorlinks=true,citecolor=blue,linkcolor=blue,urlcolor=blue]{hyperref}
\usepackage{soul}
\usepackage[normalem]{ulem}
\usepackage{dsfont}

\usepackage{algorithmicx}
\usepackage{algpseudocode}

\newcommand{\ket}[1]{\vert #1 \rangle}
\newcommand{\bra}[1]{\langle #1 \vert}

\definecolor{orcidlogocol}{HTML}{A6CE39}
\newcommand{\orcidicon}[1]{%
    \href{https://orcid.org/#1}{%
        \begin{tikzpicture}[scale=0.2, transform shape]
        \draw[orcidlogocol, fill=orcidlogocol] (0,0) circle [radius=0.5];
        \node at (0,0) {\textcolor{white}{\tiny ID}};
        \end{tikzpicture}%
    }%
}

\begin{document}

\setstcolor{red}

\title{Quantum Approximate Optimization Algorithm with Fixed Number of Parameters} 
\date{\today}

\author{Sebasti\'an Saavedra-Pino, \orcidicon{0009-0004-4229-9909}}
\affiliation{Departamento de F\'isica, CEDENNA, Universidad de Santiago de Chile (USACH), Avenida V\'ictor Jara 3493, 9170124, Santiago, Chile.}

\author{Ricardo Quispe-Mendiz\'abal, \orcidicon{0000-0001-6328-3093}}
\affiliation{Departamento de F\'isica, CEDENNA, Universidad de Santiago de Chile (USACH), Avenida V\'ictor Jara 3493, 9170124, Santiago, Chile.}

\author{Gabriel Alvarado Barrios}
\affiliation{Kipu Quantum, Greifswalderstrasse 212, 10405 Berlin, Germany}

\author{Enrique Solano, \orcidicon{0000-0002-8602-1181}}
\affiliation{Kipu Quantum, Greifswalderstrasse 212, 10405 Berlin, Germany}

\author{Juan Carlos Retamal, \orcidicon{0000-0002-7174-7879}}
\affiliation{Departamento de F\'isica, CEDENNA, Universidad de Santiago de Chile (USACH), Avenida V\'ictor Jara 3493, 9170124, Santiago, Chile.}

\author{Francisco Albarr\'an-Arriagada\,\orcidicon{0000-0001-8899-3673}}
\email[F. Albarr\'an-Arriagada]{\quad francisco.albarran@usach.cl}
\affiliation{Departamento de F\'isica, CEDENNA, Universidad de Santiago de Chile (USACH), Avenida V\'ictor Jara 3493, 9170124, Santiago, Chile.}
\affiliation{Kipu Quantum, Greifswalderstrasse 212, 10405 Berlin, Germany}

\begin{abstract}
We introduce a novel quantum optimization paradigm: the Fixed-Parameter-Count Quantum Approximate Optimization Algorithm (FPC-QAOA). It is a scalable variational framework that maintains a constant number of trainable parameters regardless of the number of qubits, Hamiltonian complexity, or circuit depth. By separating schedule function optimization from circuit digitization, FPC-QAOA enables accurate schedule approximations with minimal parameters while supporting arbitrarily deep digitized adiabatic evolutions, constrained only by NISQ hardware capabilities. This separation allows depth to scale without expanding the classical search space, mitigating overparameterization and optimization challenges typical of deep QAOA circuits, such as barren plateaus-like behaviors. We benchmark FPC-QAOA on random MaxCut instances and the Tail Assignment Problem, achieving performance comparable to or better than standard QAOA with nearly constant classical effort and significantly fewer quantum circuit evaluations. Experiments on the IBM Kingston superconducting processor with up to 50 qubits confirm robustness and hardware efficiency under realistic noise. These results position FPC-QAOA as a practical and scalable paradigm for variational quantum optimization on near-term quantum devices.
\end{abstract}

\maketitle

\section{Introduction}

Quantum computing has emerged as a transformative computational paradigm that leverages uniquely quantum resources, such as coherence and entanglement, to address problems intractable for classical computers. Over the last decade, quantum algorithms have achieved notable progress across diverse domains including chemistry, optimization, materials science, finance, and high-energy physics~\cite{Cao2019ChemRev,Phillipson2024arXiv,Weinberg2023SciRep,Dalal2024PhysRevAppl,Ma2020npjComputMat,Hofstetter2018JPB,Santagati2024NatPhys,Bauer2020ChemRev,Gisin2002RevModPhys,Pirandola2020AdvOptPhoton,Portmann2022RevModPhys,Herman2023NatRevPhys,Orus2019RevPhys,DiMeglio2024PRXQuantum}. Experimental demonstrations of quantum advantage in boson sampling~\cite{Aaronson2011Proceeding}, random circuit sampling~\cite{Boixo2018NatPhys,Arute2019Nature}, and quantum simulation~\cite{Kim2023Nature,King2025Science} highlight the growing capabilities of noisy intermediate-scale quantum (NISQ) devices~\cite{Gao2025PhysRevLett,Zhong2021PhysRevLett,Wu2021PhysRevLett,Madsen2022Nature,Acharya2025Nature}. Nevertheless, despite their significance, these demonstrations remain of limited practical relevance, motivating the development of algorithms that can address real-world optimization and simulation tasks under realistic noise and resource constraints~\cite{Chandarana2025arXiv,Simen2024PhysRevRes}.

Among the leading frameworks for near-term hardware, variational quantum algorithms (VQAs) stand out for their hybrid quantum–classical structure and shallow circuit requirements. The Quantum Approximate Optimization Algorithm (QAOA)~\cite{Farhi2014arXiv} has become a paradigmatic example, with applications in combinatorial optimization, and extensions incorporating counterdiabatic driving, Lyapunov control, and advanced schedule design~\cite{Zhou2020PRX,Blekos2024PhysRep,Chandarana2022PhysRevRes,Wurtz2022Quantum,Magann2022PhysRevA}. However, the scalability of VQAs, including QAOA, remains fundamentally limited by the rapid growth of the variational parameter space. As system size or circuit depth increases, optimization landscapes become exponentially flat, leading to barren plateaus where gradients vanish and optimization becomes infeasible~\cite{Larocca2025NatRevPhys,Wang2021NatCommun,Uvarov2021JPA,Bittel2021PhysRevLett}. This challenge imposes a fundamental barrier to scaling variational methods on NISQ hardware.

Several QAOA variants attempt to mitigate these limitations. Layerwise QAOA~\cite{Campos2021PhysRevA} constrains optimization to two parameters per iteration, alleviating gradient decay but requiring a large number of hardware evaluations and exhibiting strong sensitivity to initialization. Warm-start QAOA~\cite{Egger2021Quantum} improves the initial state through classical relaxations, yet it preserves the standard parameter scaling and remains highly dependent on the quality of the classical solver. Adaptive QAOA~\cite{Zhu2022PhysRevRes} selects operators via gradient-based criteria, reducing circuit depth while maintaining the 2p scaling and demanding extensive measurement overhead, including evaluation of the energy gradient. Although these approaches improve specific aspects of trainability or performance, none fundamentally address the dimensional growth of the variational parameter space.

In this work, we introduce the Fixed-Parameter-Count Quantum Approximate Optimization Algorithm (FPC-QAOA), a scalable modification of standard QAOA designed to separate schedule optimization from circuit digitization, breaking the dependence between circuit depth and the number of trainable parameters. In FPC-QAOA, the variational degrees of freedom remain constant regardless of the number of qubits, Trotter steps, or Hamiltonian complexity. The algorithm is formulated as a digitized adiabatic evolution governed by three smooth schedule functions associated with the initial, problem, and auxiliary Hamiltonians. These schedules are parametrized by a fixed number of parameters, enabling deep and expressive ansätze without expanding the optimization space. Unlike standard QAOA, which intertwines digitization and schedule optimization—requiring large parameter sets for accurate discretization—our framework allows arbitrarily deep digitalizations limited only by hardware capabilities, without increasing classical optimization complexity. By avoiding overparameterization, FPC-QAOA mitigates barren-plateau-like behaviors, stabilizes convergence, and reduces classical effort. Consequently, the method requires significantly fewer quantum circuit evaluations, lowering processor usage and operational cost. These features establish FPC-QAOA as a practical and hardware-efficient framework for scalable quantum optimization on near-term quantum devices.

\section{FPC-QAOA}

\begin{figure*}[t!]
  \centering
  \includegraphics[width=0.9\linewidth]{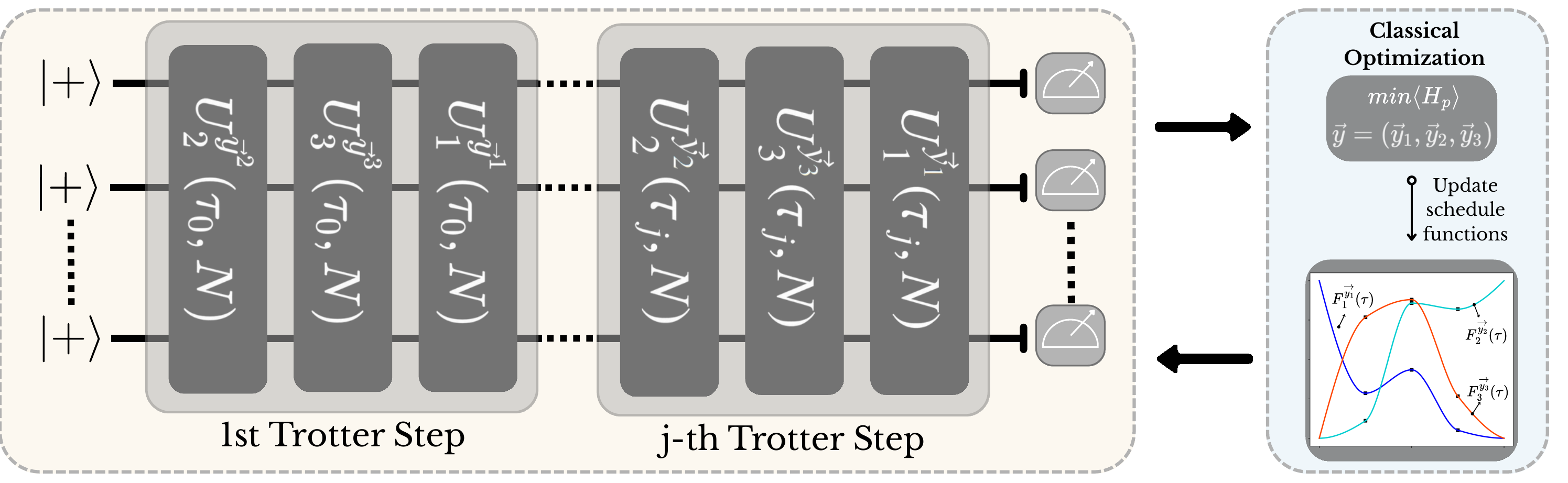}
  \caption{
  Conceptual overview of the Fixed-Parameter-Count Quantum Approximate Optimization Algorithm (FPC-QAOA).
  Three trainable, monotone schedule functions $F_1$, $F_2$, and $F_3$ are reconstructed via cubic Hermite interpolation from a fixed set of internal control points.
  At each Trotter step $j$, the schedules are sampled at time $\tau_j$ to generate angle parameters associated with the initial Hamiltonian $H_i$ ($R_x$ rotations), the problem Hamiltonian $H_p$ ($R_z$ and $R_{zz}$ evolutions), and the auxiliary Hamiltonian $H_{\mathrm{aux}}$ (local $R_z$ rotations).
  The total number of trainable parameters is independent of the number of Trotter steps $N$, allowing increasingly accurate digitalization without increasing the dimensionality of the classical optimization problem.
  }
  \label{Fig01}
\end{figure*}

QAOA and adiabatic quantum computing are closely related, as both aim to prepare the ground state of a target Hamiltonian through an analog or digitized adiabatic evolution. In adiabatic quantum computing, the system evolves from the ground state of an initial Hamiltonian $H_i$ to that of a problem Hamiltonian $H_p$ under the time-dependent Hamiltonian
\begin{equation}
    H(t) = [1 - \lambda(t/T)] H_i + \lambda(t/T) H_p,
    \label{Eq01}
\end{equation}
where $\lambda(t/T)$ is a scheduling function satisfying $\lambda(0)=0$ and $\lambda(1)=1$.
If the total evolution time $T$ is sufficiently long, the adiabatic theorem guarantees that the system remains close to its instantaneous ground state,
\begin{equation}
    \ket{\psi(t)} =
    \mathcal{T}
    \exp\!\left(-\frac{i}{\hbar}\int_0^t H(\tau)\, d\tau\right)
    \ket{\psi(0)},
    \label{Eq02}
\end{equation}
where $\mathcal{T}$ denotes time ordering.

By discretizing time and applying a Trotter expansion, the evolution can be approximated as
\begin{eqnarray}
    \ket{\psi(t)} &\approx&
    \prod_{j} e^{-\frac{i}{\hbar}H(t_j)\Delta t_j}\ket{\psi(0)} \nonumber\\
    &=&
    \prod_{j}
    e^{-\frac{i}{\hbar}[1-\lambda(t_j/T)]H_i\Delta t_j}
    e^{-\frac{i}{\hbar}\lambda(t_j/T)H_p\Delta t_j}
    \ket{\psi(0)},\nonumber\\
    \label{Eq03}
\end{eqnarray}
with $\Delta t_j = t_{j+1}-t_j$.
For $N$ Trotter steps, defining
$\alpha_j = [1-\lambda(t_j/T)]\Delta t_j$ and
$\beta_j = \lambda(t_j/T)\Delta t_j$,
the state reads
\begin{equation}
    \ket{\psi(t)}_{QAOA} \approx
    \prod_{j=1}^{N}
    e^{-i\alpha_j H_i}
    e^{-i\beta_j H_p}
    \ket{\psi(0)}.
    \label{Eq04}
\end{equation}
For the common choice $H_i = -\epsilon\sum_j \sigma_j^x$, the initial state is
$\ket{\psi(0)}=\ket{++\cdots+}$, and optimizing the parameters
$\{\alpha_j,\beta_j\}$ recovers the standard QAOA.
Thus, QAOA can be interpreted as a variationally optimized, digitized adiabatic evolution with $2N$ variational parameters.

While increasing the number of Trotter steps improves the accuracy of the digital approximation, due to the linear growth of the variational parameter space, the classical optimization routine becomes harder and more complex. This scaling behavior constitutes a fundamental problem to the application of QAOA to large-scale or industrially relevant problem instances.

\subsection{Fixed-parameter-count formulation}

To overcome this limitation, we now describe the framework of our proposed Fixed-Parameter-Count Quantum Approximate Optimization Algorithm (FPC-QAOA), whose overall workflow is schematically illustrated in Fig.~\ref{Fig01}.
The goal of FPC-QAOA is to generalize the adiabatic evolution while fixing the number of trainable parameters, independently of the circuit depth.

We introduce a generalized time-dependent Hamiltonian of the form
\begin{equation}
    H(t) = F_1(t/T) H_i + F_2(t/T) H_p + F_3(t/T) H_{\mathrm{aux}},
    \label{Eq05}
\end{equation}
where $F_1$, $F_2$, and $F_3$ are smooth schedule functions satisfying the boundary conditions
$F_1(0)=F_2(1)=1$ and
$F_1(1)=F_2(0)=F_3(0)=F_3(1)=0$.
The auxiliary Hamiltonian
\begin{equation}
    H_{\mathrm{aux}} = \sum_{j=1}^{n}\omega_j \sigma_j^z
\end{equation}
corresponds to the local component of the problem Hamiltonian, where $n$ denotes the number of qubits.
This term acts as a time-dependent local bias, which can modify  the instantaneous spectrum helping to avoid detrimental small gaps or level crossings.

\subsection{Schedule parameterization with fixed complexity}

Each schedule function $F_k(s)$, with $s=t/T\in[0,1]$, is parameterized by a fixed number of internal control points in addition to the boundary conditions.
Let $n_p$ denote the number of variational parameters per schedule.
We define a uniform grid
$\vec{s}=\{s_0,s_1,\ldots,s_{n_p},s_{n_p+1}\}$ with
$s_0=0$,
$s_j=\frac{j}{n_p+1}$,
and $s_{n_p+1}=1$,
together with the parameter set
$\boldsymbol{y}^{k}=\{y^{k}_1,\dots,y^{k}_{n_p}\}$ for the $k$th schedule.
This construction defines the point set
$\mathcal{P}_k=\{(s_0,F_k(0)),\ldots,(s_j,y_j^{k}),\ldots,(s_{n_p+1},F_k(1))\}$.

Each schedule function $F_k(s)$ is reconstructed by interpolating $\mathcal{P}_k$ using cubic Hermite interpolation, following Refs.~\cite{Barraza2022IOP,barraza2023arXiv}.
This procedure ensures that $F_k(s)$ is smooth, continuous, and monotonic between adjacent control points.

In the digital implementation with $N$ Trotter steps, the evolution operators are evaluated at the midpoints (second-order accurate discretization) 
\begin{equation}
    \tau_j = \frac{T}{2N} + j\frac{T}{N},
\end{equation}
leading to the approximation
\begin{equation}
    \ket{\psi(T)}_{FPC} \approx
    \prod_{j=1}^{N}
    U^{\boldsymbol{y}_1}_1(\tau_j,N)
    U^{\boldsymbol{y}_2}_2(\tau_j,N)
    U^{\boldsymbol{y}_3}_3(\tau_j,N)
    \ket{\psi(0)},
    \label{Eq07}
\end{equation}
where
\begin{equation}
    U^{\boldsymbol{y}_k}_k(\tau_j,N)
    =
    \exp\!\left[
    -\frac{i}{\hbar}
    F^{\boldsymbol{y}_k}_k(\tau_j/T)
    \frac{T}{N}
    H_k
    \right],
\end{equation}
with $H_1=H_i$, $H_2=H_p$, and $H_3=H_{\mathrm{aux}}$.
Accordingly, $U^{\boldsymbol{y}_1}_1$ implements local $R_x$ rotations,
$U^{\boldsymbol{y}_2}_2$ corresponds to $R_z$ and $R_{zz}$ evolutions for QUBO Hamiltonians,
and $U^{\boldsymbol{y}_3}_3$ yields local $R_z$ rotations.
It is important to highlight that the total number of trainable parameters ($3n_p$) is independent of the number of Trotter steps $N$, enabling deeper digitalization (smaller Trotter error), without increasing the dimensionality of the classical optimization.

\subsection{Cost function and performance metrics}

For both QAOA and FPC-QAOA, the cost function is defined as the expectation value of the problem Hamiltonian $H_p$ evaluated at the end of the evolution,
\begin{equation}
    E_{abc}
    =
    \bra{\psi_{abc}(T)} H_p \ket{\psi_{abc}(T)},
\end{equation}
where $\ket{\psi_{abc}(T)}$ denotes the final state obtained from Eq.~(\ref{Eq04}) for QAOA or Eq.~(\ref{Eq07}) for FPC-QAOA, respectively.
Classical optimization is performed using the gradient-free COBYLA algorithm in combination with the Conditional Value-at-Risk (CVaR) objective~\cite{Barkoutsos2020Quantum}, which enhances robustness and convergence.

Performance is quantified through the normalized energy reduction,
\begin{equation}
    \mathcal{R}_{abc} =
    \frac{\bra{\psi(0)}H_p\ket{\psi(0)}
    - E_{abc}}
    {\bra{\psi(0)}H_p\ket{\psi(0)}-E_0},
    \label{Eq09}
\end{equation}
where $0 \le \mathcal{R}_{abc} \le 1$.
To directly compare both algorithms, we define the enhancement ratio
\begin{equation}
    \eta =
    \frac{\mathcal{R}_{\mathrm{FPC\text{-}QAOA}}}
         {\mathcal{R}_{\mathrm{QAOA}}},
    \label{Eq10}
\end{equation}
with $\eta>1$ indicating superior performance of FPC-QAOA.

\section{Results}

\subsection{Random MaxCut graphs}

\begin{figure}[t!]
  \centering
  \includegraphics[width=0.9\linewidth]{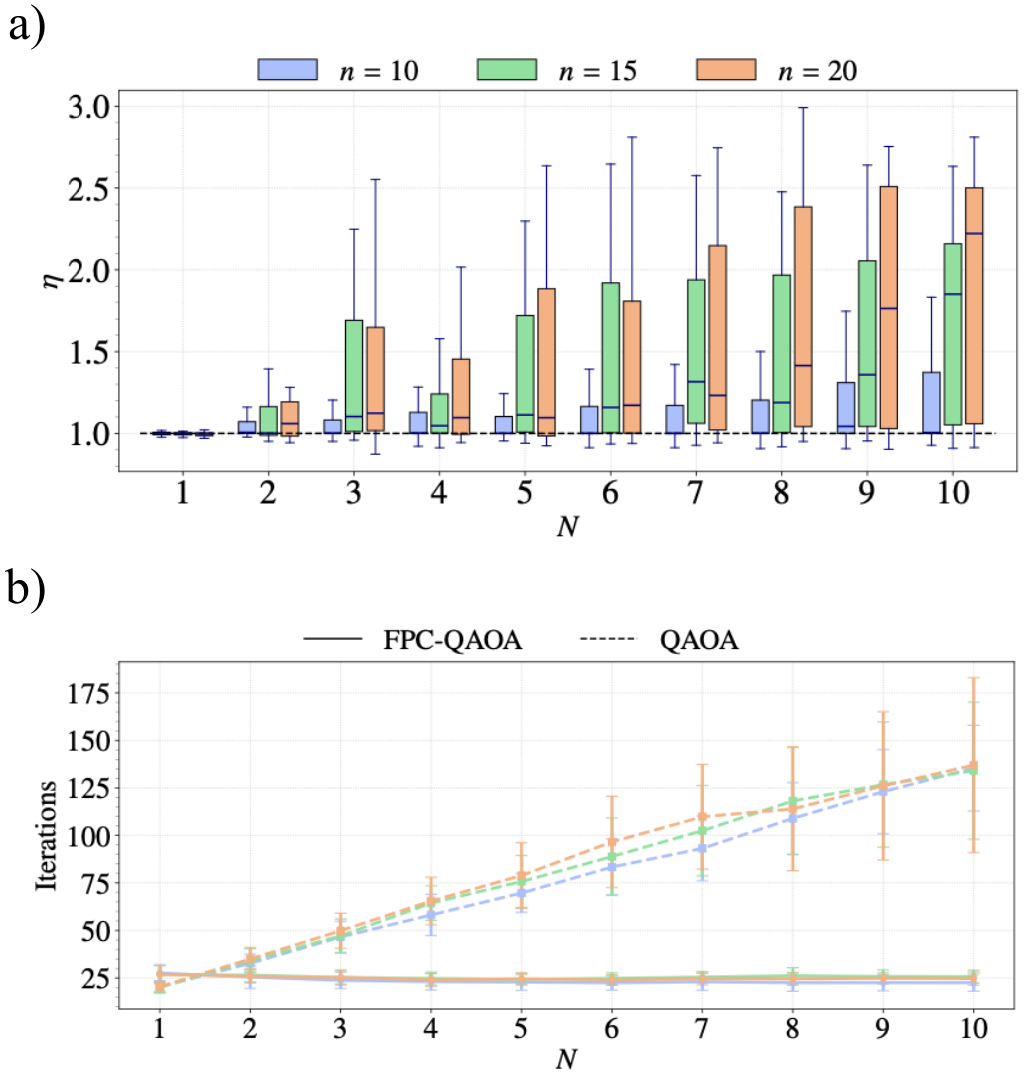}
  \caption{(a) Enhancement ratio $\eta$ over 100 random MaxCut instances (up to $n=20$ nodes/qubits) for FPC-QAOA with three trainable parameters (one per schedule). Medians exceed one, indicating consistent improvement over standard QAOA.
    (b) Average number of classical optimization iterations versus Trotter depth. While QAOA requires an increasing number of iterations as depth grows, FPC-QAOA remains nearly constant, consistent with a fixed-size parameterization across depths.}
  \label{Fig02}
\end{figure}

We first benchmark FPC-QAOA on the MaxCut problem~\cite{Glover2022AOR}.
Given a graph $G(V,E)$ with vertex set $V$ and edge set $E$, the objective is to partition $V$ into two subsets maximizing the number of crossing edges,
\begin{equation}
\text{maximize}\;\; y=\sum_{(j,k)\in E}\left(x_j+x_k-2x_jx_k\right),
\label{Eq11}
\end{equation}
with $x_j\in\{0,1\}$.
Promoting binary variables to spin operators $x_j\to (1-\sigma^z_j)/2$ and flipping the sign, we obtain an Ising-type Hamiltonian (up to an additive constant) for minimization,
\begin{equation}
H_p=\frac{1}{2}\sum_{(j,k)\in E}\left(\sigma^z_j\sigma^z_k-I\right).
\label{Eq12}
\end{equation}

We compare QAOA and FPC-QAOA on random $k$-regular graphs with $k=n/2$ and $n\le 20$ (number of nodes/qubits), using IBM Aer with $10{,}000$ shots per circuit evaluation.
Figure~\ref{Fig02}(a) reports the enhancement ratio $\eta$ (see Eq.~(\ref{Eq10})), showing $\eta>1$ for most instances while using only three trainable parameters in FPC-QAOA.
Figure~\ref{Fig02}(b) shows that the number of classical optimization iterations required by FPC-QAOA is nearly independent of depth, in contrast to standard QAOA, indicating that fixing the parameter count stabilizes the optimization effort as the Trotter depth increases.

\begin{figure}[t]
    \centering
    \includegraphics[width=\linewidth]{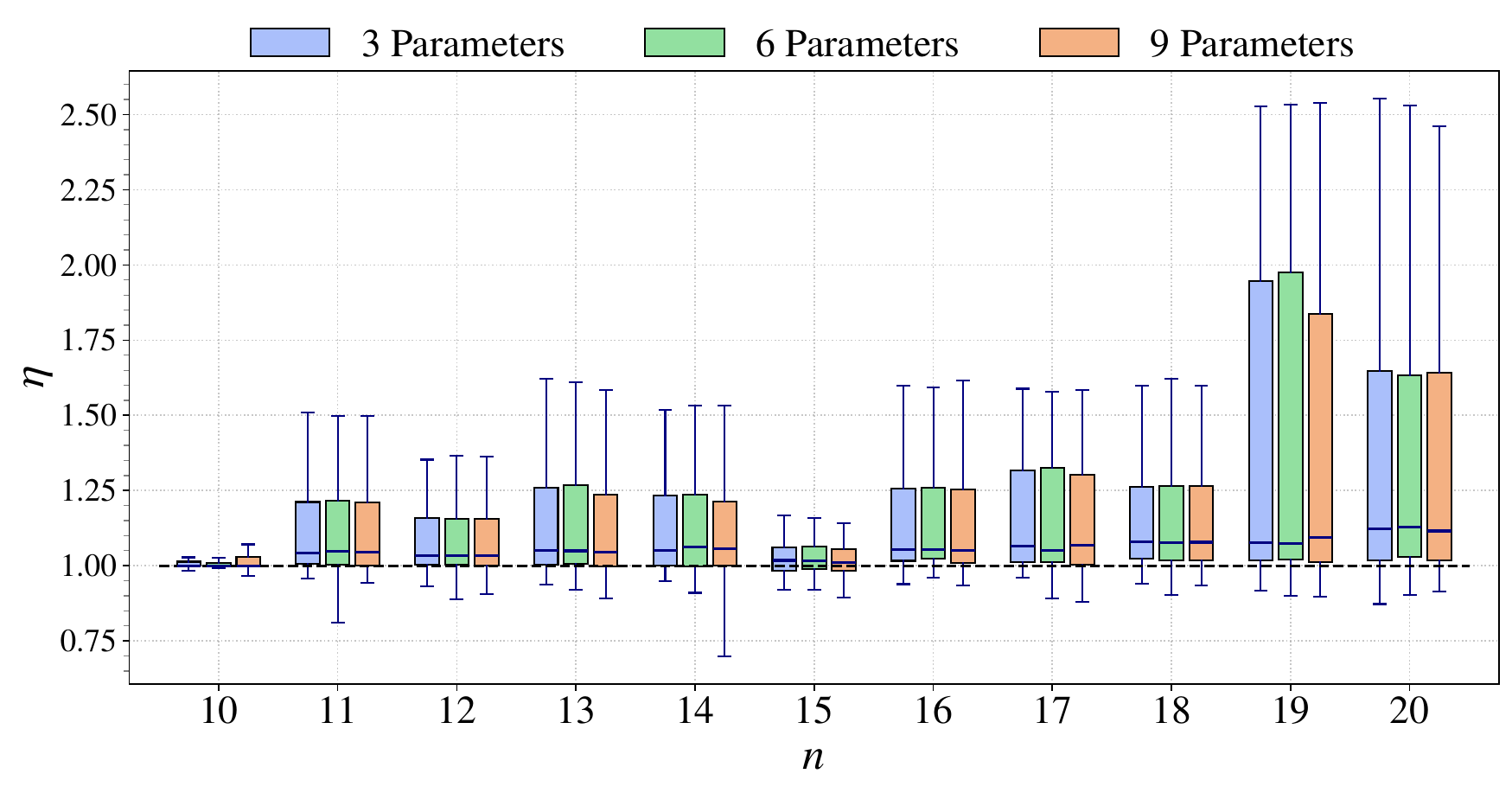}
    \caption{
    Enhancement ratio $\eta$ versus system size ($n=10$--$20$ nodes) at fixed Trotter depth ($N=3$).
    Colors indicate the total number of trainable parameters used by FPC-QAOA (3, 6, 9) and the corresponding QAOA baseline (6 parameters at three layers).
    FPC-QAOA maintains $\eta\gtrsim 1$ across sizes, with gains that tend to increase as the problem size grows.
    }
    \label{Fig03}
\end{figure}

Figure~\ref{Fig03} summarizes scaling with system size by increasing the number of nodes $n$ at fixed Trotter depth ($N=3$).
FPC-QAOA matches or outperforms QAOA across all tested sizes, with improvements that tend to grow with $n$.
This trend is consistent with the fact that larger instances typically require deeper standard-QAOA circuits (and therefore more parameters) to maintain solution quality, whereas FPC-QAOA can increase depth without expanding the optimization space.

\subsection{$(C,S,W)_n$ connectivity}

\begin{figure}[b]
    \centering
    \includegraphics[width=0.8\linewidth]{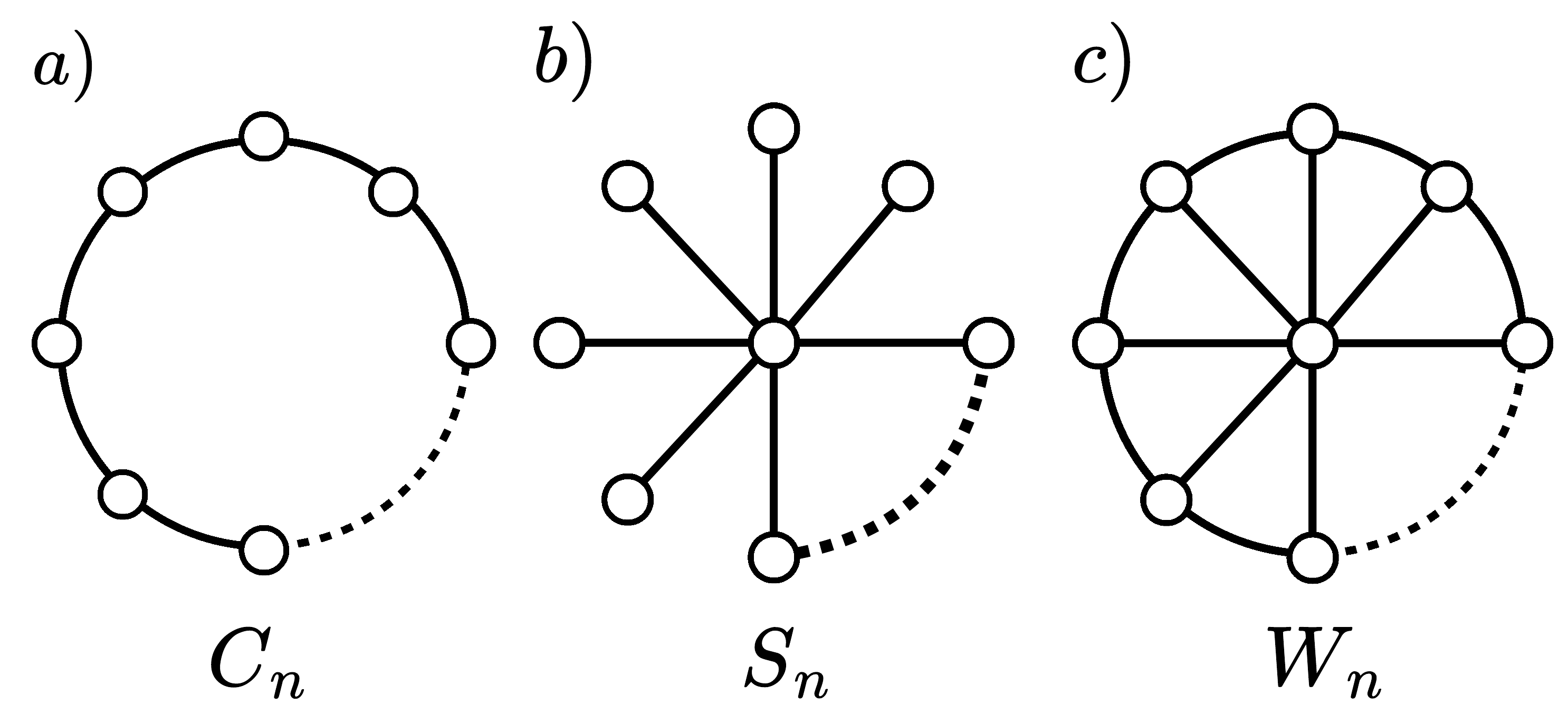}
    \caption{
    Graph topologies considered in the connectivity study:
    (a) cyclic ($C_n$) topology with periodic boundary conditions,
    (b) star ($S_n$) topology, and
    (c) wheel ($W_n$) topology combining cycle and star.
    }
    \label{Fig04}
\end{figure}

To probe topology-dependent behavior, we study the three structures shown in Fig.~\ref{Fig04}:
(i) cyclic connectivity ($C_n$),
(ii) star connectivity ($S_n$), and
(iii) wheel connectivity ($W_n$).
We consider the corresponding Ising Hamiltonians
\begin{equation}
H_C=\sum_{j=1}^{n}\omega_j\sigma^z_j+\sum_{j=1}^{n}g_j\,\sigma^z_j\sigma^z_{j+1},
\quad (n{+}1)\equiv 1,
\end{equation}
\begin{equation}
H_S=\sum_{j=1}^{n}\omega_j\sigma^z_j+\sum_{j=2}^{n}g_j\,\sigma^z_j\sigma^z_{1},
\end{equation}
\begin{equation}
H_W=\sum_{j=1}^{n}\omega_j\sigma^z_j+\sum_{j=2}^{n}g_j\,\sigma^z_j\sigma^z_{1}
+\sum_{j=2}^{n}h_j\,\sigma^z_j\sigma^z_{j+1},
\quad (n{+}1)\equiv 2 .
\end{equation}

\begin{figure}[b]
    \centering
    \includegraphics[width=\linewidth]{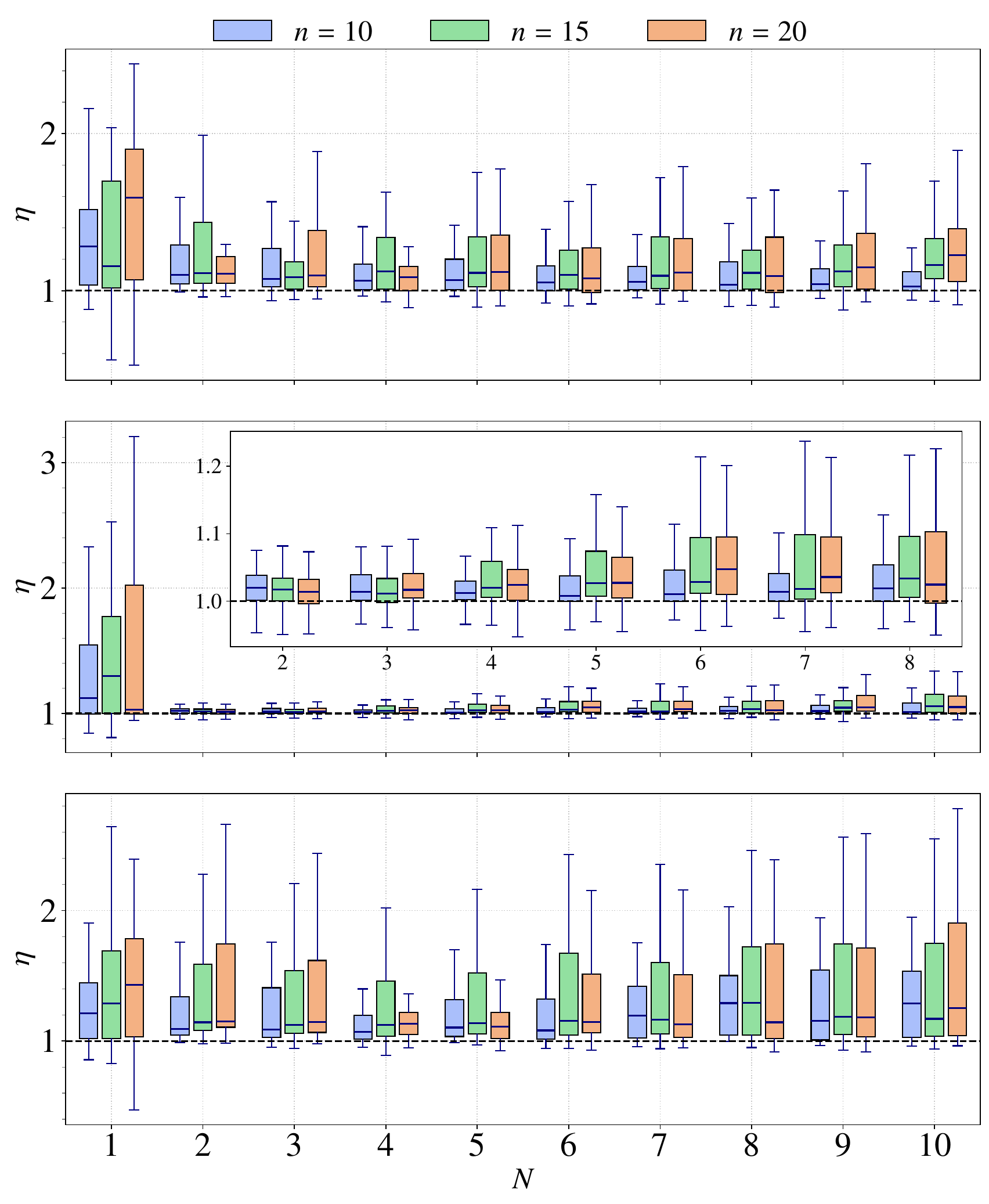}
    \caption{
    Enhancement ratio $\eta$ for $C_n$ (top), $S_n$ (middle), and $W_n$ (bottom).
    Each panel aggregates 100 random instances at $n\in\{10,15,20\}$ with local and pairwise weights sampled uniformly in $[-1,1]$.
    FPC-QAOA uses three trainable parameters, whereas the QAOA baseline uses $2N$ parameters at the same depth.
    Medians above one indicate consistent gains, with smaller---but still positive---improvements on star graphs.
    }
    \label{Fig05}
\end{figure}

For each topology, we consider $n\in\{10,15,20\}$ and generate 100 random instances with weights sampled uniformly in $[-1,1]$.
All results are obtained with IBM Aer using $10{,}000$ shots per instance.
As shown in Fig.~\ref{Fig05}, FPC-QAOA consistently improves over QAOA across topologies.
The star topology exhibits smaller gains, whereas cycle and wheel graphs show stronger enhancement.
Importantly, these improvements are obtained with only three trainable parameters, which translates into fewer classical iterations and fewer total circuit evaluations, consistent with the MaxCut benchmark above.

\subsection{Tail-Assignment Problem}

We next evaluate FPC-QAOA on the Tail-Assignment Problem (TAP), a logistics optimization task in which available routes $r\in\mathcal{R}$ are assigned to flights $f\in\mathcal{F}$ to satisfy coverage constraints at minimum operational cost~\cite{Vikstal2020PhysRevAppl,Gili2024arXiv}.
Let $x_r\in\{0,1\}$ denote route selection, $c_r$ its cost, and $A=\{a_{f,r}\}$ the incidence matrix, where $a_{f,r}\in\{0,1\}$ indicates whether route $r$ covers flight $f$.
Enforcing that every flight is covered, $\sum_{r\in\mathcal{R}} a_{f,r} x_r \ge 1$ for all $f$, yields the quadratic cost function
\begin{equation}
Q(x)=\underbrace{\sum_{r\in\mathcal{R}}c_r x_r}_{\text{Operational cost}}
+ \underbrace{P\sum_{f\in\mathcal{F}}\left(1-\sum_{r\in\mathcal{R}}a_{f,r}x_r\right)^2}_{\text{Penalty}}.
\end{equation}
Mapping $x_r \mapsto (1-\hat{\sigma}^{r}_z)/2$ and neglecting constants, we obtain the Ising form
\begin{equation}
H_{\mathrm{TAP}}=\sum_r h_r\,\hat\sigma^r_z+\sum_{r<r'}J_{rr'}\,\hat\sigma^r_z\hat\sigma^{r'}_z,
\end{equation}
with coefficients
\begin{equation}
\begin{aligned}
h_r &=-\frac{1}{2}c_r-\frac{P}{2}\sum_f a_{fr}\left(\sum_{r'} a_{fr'}-2\right),\\
J_{rr'} &=\frac{P}{4}\sum_f a_{fr}a_{fr'}.
\end{aligned}
\end{equation}

We consider $n_r\in\{10,15,20\}$ routes and $n_f=5n_r$ flights.
For each configuration, we generate 100 random instances with $A\in\{0,1\}^{n_f\times n_r}$ and costs $c_r\in[2,10]$. All simulations use $10{,}000$ shots per instance in a noiseless backend.

\begin{figure}[t]
    \centering
    \includegraphics[width=\linewidth]{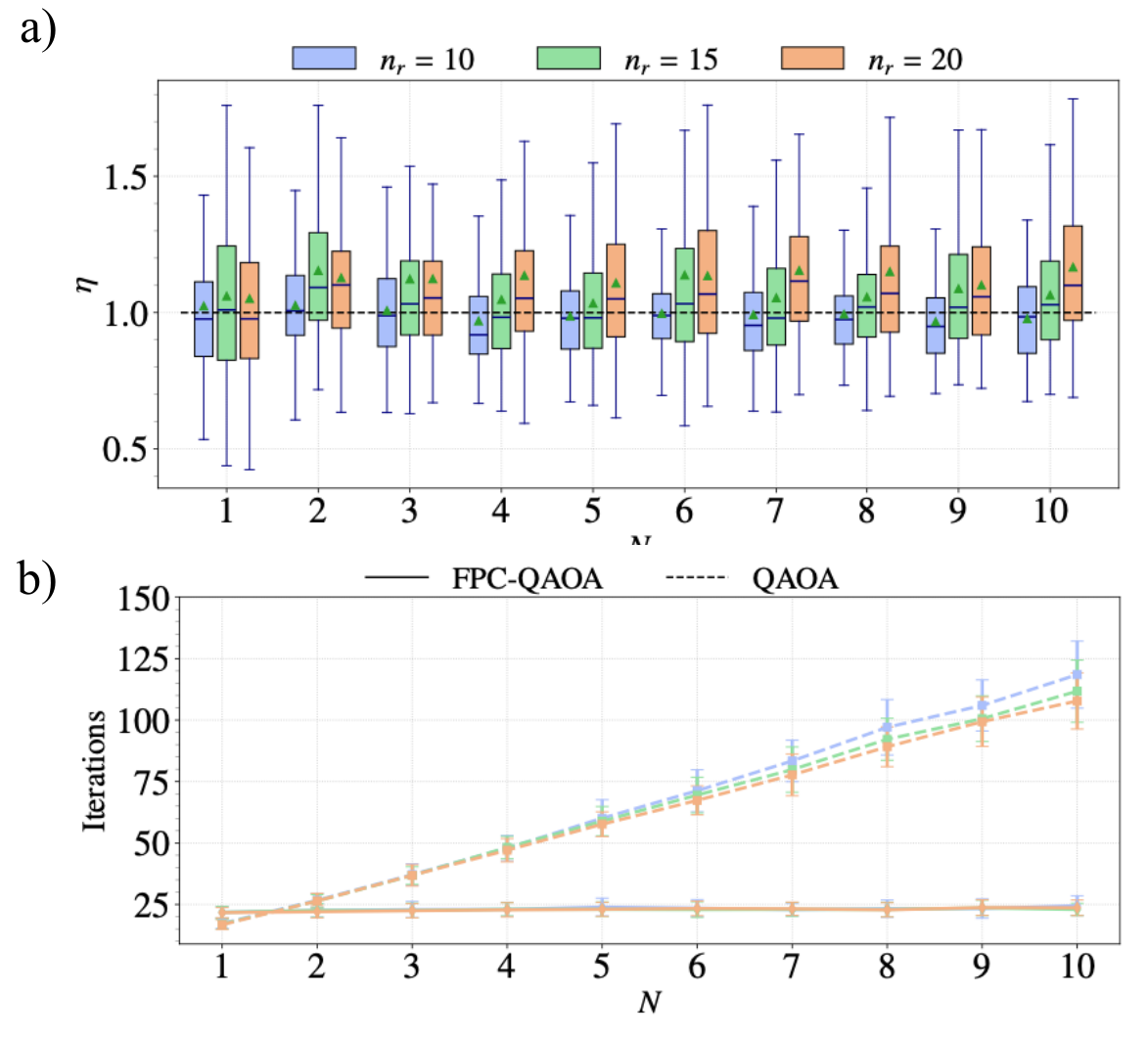}
    \caption{
    (a) Enhancement ratio $\eta$ across 100 TAP instances for $n_r\in\{10,15,20\}$ and varying Trotter depth $N$.
    Lines mark medians and triangles mark means.
    (b) Average number of classical optimization iterations versus depth.
    }
    \label{Fig06}
\end{figure}

Figure~\ref{Fig06} shows that the mean enhancement ratio is slightly above one across problem sizes, while the number of classical iterations required by FPC-QAOA remains nearly constant with depth.
In contrast, QAOA exhibits an approximately linear growth in the number of iterations as depth increases.
This behavior indicates that fixing the parameter count reduces optimization effort and, consequently, the total number of circuit evaluations required to reach convergence.

\subsection{IBM quantum hardware results}
\label{subsec:ibm_hw}

Finally, we implement FPC-QAOA on IBM superconducting quantum hardware and compare its performance against standard QAOA and random sampling.
Random sampling draws a fixed number of candidate bit strings (shots) uniformly at random, and the best observed solution serves as a baseline.
While simple, its effectiveness decays rapidly with problem size due to the exponential growth of the search space.

To ensure a fair comparison, we use the same number of Trotter steps, the same number of shots ($10{,}000$), identical transpilation constraints, and identical CVaR post-processing for QAOA and FPC-QAOA.
TAP instances are generated with $n_r=50$ routes and $n_f=180$ flights, with an average node valency of $2.7$.
We run both algorithms at depths $N\in\{3,5\}$.
All circuits are transpiled for the \texttt{ibm\_kingston} backend subject to its connectivity and calibration constraints, and we enforce comparable circuit depths for QAOA and FPC-QAOA (not exceeding $\sim 600$ layers) to remain within device coherence limits.
Classical optimization uses COBYLA with the same stopping criteria in all cases.

\begin{figure}[t]
    \centering
    \includegraphics[width=\linewidth]{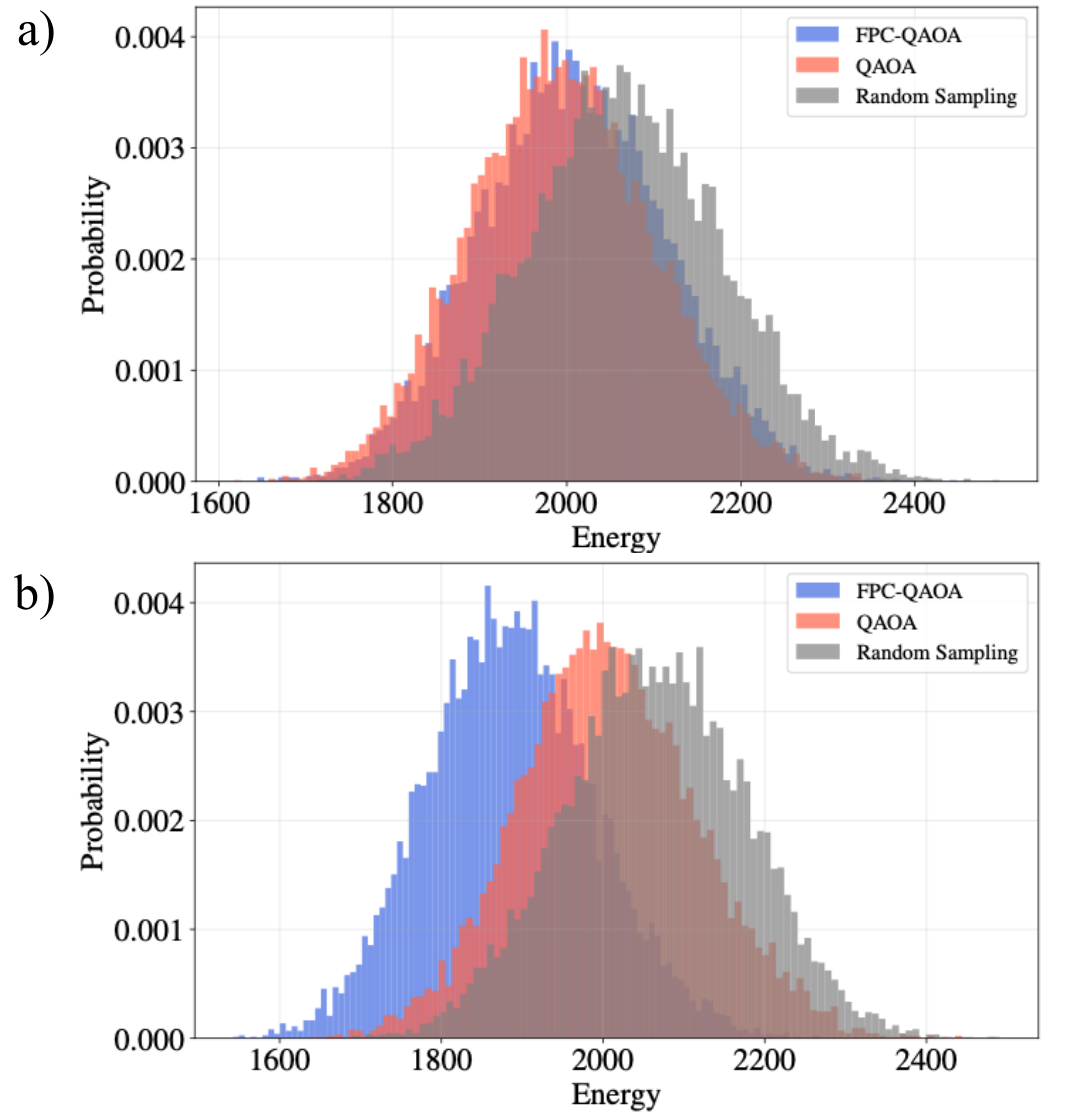}
    \caption{
    \texttt{ibm\_kingston} hardware results for the TAP instance: measured energy distributions for FPC-QAOA, QAOA, and random sampling.
    (a) Experiments executed with $N=3$ Trotter steps.
    (b) Experiments executed with $N=5$ Trotter steps.
    Each experiment considers a maximum of 25 function evaluations in the classical optimizer.
    }
    \label{Fig07}
\end{figure}

Figure~\ref{Fig07} reports the measured energy distributions on hardware, and Table~\ref{tab:results_hw_tap} summarizes the average energy, best observed energy, and the number of optimizer iterations in each case.
Consistent with the simulator benchmarks, the relative advantage of FPC-QAOA becomes more pronounced at larger Trotter depth.
Moreover, across depths FPC-QAOA attains competitive energies while typically requiring fewer optimization iterations, consistent with the fixed-parameter-count design that avoids an increasing number of variational degrees of freedom as depth grows.

\begin{table*}[t]
\centering
\caption{Performance comparison of FPC-QAOA and QAOA on the hardware TAP instance.}
\label{tab:results_hw_tap}

\renewcommand{\arraystretch}{1.2}
\setlength{\tabcolsep}{10pt}

\begin{tabular}{|l|c|c|c|c|}
\hline
\multicolumn{5}{|c|}{\textbf{TAIL-ASSIGNMENT (hardware instance)}} \\ \hline
\textbf{Algorithm} & \textbf{$N$} & \textbf{Avg. Energy} & \textbf{Best Energy} & \textbf{Iterations} \\  \hline
FPC-QAOA & 3 & 1857.58 & 1644.98 & 19  \\ 
QAOA     & 3 & 1843.98 & 1617.34 & 25  \\ 
FPC-QAOA & 5 & 1741.93 & 1542.41 & 20  \\ 
QAOA     & 5 & 1853.48 & 1634.56 & 25  \\ \hline
\end{tabular}
\end{table*}

\section{Analysis and Conclusion}

In this work, we have introduced the Fixed-Parameter-Count Quantum Approximate Optimization Algorithm (FPC-QAOA), a scalable framework that fundamentally addresses one of the main limitations of variational quantum algorithms: the exponential growth of the parameter space with circuit depth. By separating schedule optimization from circuit digitization, FPC-QAOA enables arbitrarily deep digitized adiabatic evolutions constrained only by hardware capabilities, without increasing the number of trainable parameters or the complexity of the classical optimization process. This design mitigates barren-plateau-like behaviors, stabilizes convergence, and reduces overparameterization, allowing expressive ansätze with minimal classical overhead.

FPC-QAOA preserves the core structure of QAOA while recasting the ansatz as a digitized adiabatic evolution governed by a small set of smooth schedule functions.
In particular, three monotone schedules associated with the initial, problem, and auxiliary Hamiltonians define a compact parameterization that remains fixed as the depth increases. This fixed-size parameterization enables deeper digitalization without expanding the classical search space, thereby reducing the sensitivity of training to depth-dependent overparameterization.

Across numerical benchmarks, FPC-QAOA exhibited systematic improvements over standard QAOA in both solution quality and optimization effort. For random MaxCut instances, the enhancement ratio $\eta$ was typically above one, indicating improved normalized energy reduction relative to QAOA at matched depth. In addition, while the number of COBYLA iterations required by QAOA increased with depth, FPC-QAOA maintained nearly depth-independent iteration counts, consistent with a more stable optimization process when the parameter count is fixed.

We further assessed topology-dependent behavior using cyclic ($C_n$), star ($S_n$), and wheel ($W_n$) connectivity.
FPC-QAOA consistently matched or improved upon QAOA while using significantly fewer parameters.
Although gains were less pronounced for the star topology, the method remained competitive and retained its advantage: performance improvements obtained without increasing the dimensionality of the problem.

To evaluate performance on an industry-motivated task, we applied FPC-QAOA to the Tail-Assignment Problem (TAP). Across instance sizes and depths, FPC-QAOA achieved competitive or improved energies compared to QAOA, while maintaining stable convergence behavior with a constant parameter count. These results support the relevance of fixed-parameter-count designs for near-term devices, where classical optimization cost and measurement budgets are critical constraints.

A further operational advantage of FPC-QAOA is the reduction in quantum-hardware usage implied by faster convergence. Fewer optimizer iterations translate into fewer objective-function evaluations and, consequently, fewer circuit executions at a fixed shot budget. This directly reduces quantum processor time and can lower both operational cost and queue time in access-limited NISQ environments.

Several directions merit future investigation. First, extending fixed-parameter-count schedule parameterizations beyond QUBO/Ising formulations to broader classes of optimization and simulation problems remains an important goal. Second, it will be valuable to study alternative auxiliary terms and schedule constraints tailored to specific problem structures, as well as robustness under realistic noise models and hardware-aware compilation. Finally, adapting the fixed-parameter-count paradigm to other variational algorithms may provide a general route to improving scalability and trainability on near-term hardware.

In summary, FPC-QAOA provides a hardware-efficient, scalable framework for variational quantum optimization by enabling deeper circuits without increasing the number of trainable parameters. Our results indicate that this design can improve solution quality and reduce optimization effort across benchmarks and on representative hardware experiments, supporting its potential as a practical approach for near-term quantum optimization.

\section{Data availability}
Data supporting our findings are available from the corresponding author, upon reasonable request.

\section{Acknowledgments}

We acknowledge financial support from Agencia Nacional de Investigaci\'on y Desarrollo (ANID): Financiamiento Basal para Centros Cient\'ificos y Tecnol\'ogicos de Excelencia grant No. CIA250002, Fondecyt grant 1231172 and Fondecyt Exploraci\'on grant No. 13250014. Also the financial support from Universidad de Santiago de Chile: DICYT Asociativo Grant No. 042431AA\_DAS.
\newpage
\appendix

\section{FPC-QAOA pseudocode}
\label{AppA}

\begin{figure}[h]
\caption{schedule\_functions}
\label{schedule_functions}
\begin{minipage}{\linewidth}
\footnotesize
\begin{algorithmic}[1]
    \State P: Random numbers array with length $3p$ with $p$ the number of parameters for each schedule function.
    \State $F_i$: i-th Parametrized scheduling function.
    \State $P_i$: Parameters assigned from the array P to the i-th schedule function.
    \State $f_i$: i-th Array with corresponding $P_i$ and boundary conditions.
    \State Interpolator\_function: equal-spaced point interpolator.
    \State t: Array of normalized times from 0 to 1, with length $3p + 2$
    \State
    \State from P assign $P_1, P_2 \text{ and }P_3$
    \State Assert that each $P_i$ has the same length
    \State $F_1$ = Interpolation between t and $f_1$
    \State $F_2$ = Interpolation between t and $f_2$
    \State $F_3$ = Interpolation between t and $f_3$
    \State Save interpolation objects
\State \textbf{end}
\end{algorithmic}
\end{minipage}
\end{figure}

\begin{figure}[t]
\caption{fpcqaoa\_ansatz}
\label{FVP-QAOA}
\begin{minipage}{\linewidth}
\footnotesize
\begin{algorithmic}[1]
    \State Parameters: Random numbers array with length 3p with p number of parameters for each schedule function.
    \State $F_i$: i-th Parametrized-interpolated scheduling function.
    \State $\epsilon$: $H_i$ Frequency
    \State T: Final time of the adiabatic evolution
    \State layers: Number of Trotter Steps   
    \State n: Number of qubits
    \State Rx: Local $\sigma_x$ interaction
    \State Rz: Local $\sigma_z$ interaction
    \State Rzz: Bi-local $\sigma_z$ interaction
    \State H: Hadamard Gate
    \State edges: Data set with the nodes interaction and weights of the graphs
    \State
    \State $\Delta t$ = T/layers
    \State ($F_1$, $F_2$, $F_3$) = \textbf{schedule\_functions(Parameters)}
    \State
    \State Create a new quantum circuit
    \For{k1 from 0 to n-1}
        \State Apply H to the k1-th qubit
        \EndFor
    \For{k2 from 0 to layers-1}
        \State times\_sf = (k2 + 0.5) $\Delta t$ / T
        \For{each node interactions and weights in edges}
            \If{no nodes interactions in edges}
            \State Continue
            \ElsIf{only local interactions in the node}
            \State $\theta_f = 2\text{weights}\Delta t F_2(\text{times\_sf})$
            \State Apply Rz to with angle $\theta_f$ for the local interaction
            \ElsIf{only bi-local interactions in the nodes}
            \State $\theta_f = 2\text{weights}\Delta t F_2(\text{times\_sf})$
            \State Apply Rzz to with angle $\theta_f$ for the bi-local interaction
            \EndIf
            \EndFor
            
    \For{k3 from 0 to n-1}
        \State $\theta_a = 2\text{weights\_aux} \Delta t F_3(\text{times\_sf)}$
        \State Apply Rz with angle $\theta_a$ to the k3-th qubit)
        \EndFor

    \For{k4 from 0 to n-1}
        \State $\theta_i = -2 \epsilon \Delta t F_1(\text{times\_sf)}$
        \State Apply Rx with angle $\theta_i$ to the k4-th qubits
        \EndFor

    \EndFor
\State Measure All Active Qubits
    
\State \textbf{end}
\end{algorithmic}
\end{minipage}
\end{figure}

\newpage

\input{FVP-QAOA.bbl}
\end{document}

%% file: FVP-QAOA.bbl
\providecommand{\noopsort}[1]{}\providecommand{\singleletter}[1]{#1}%